\documentclass[pre,aps,twocolumn,nofootinbib]{revtex4-1}

\pdfoutput=1

\usepackage[T1]{fontenc}
\usepackage{amsmath}
\usepackage{graphicx}
\usepackage{subfig}
\usepackage{ulem}
\usepackage{amssymb,booktabs}
\usepackage[arrow, matrix, curve]{xy}
\usepackage{xcolor}
\usepackage{dcolumn}
\usepackage{bm}
\usepackage{ulem}

\usepackage{color}

\newcounter{comment}
\newcommand{\dd}{\mathrm d}

\begin{document}


\title{Phenomenological modeling of the motility of self-propelled microorganisms}

\author{Silvia Zaoli $^{1}$, Andrea Giometto$^{2,3}$, Marco Formentin$^{1}$, Sandro Azaele$^{4}$,   Andrea Rinaldo$^{2,5}$,  Amos Maritan$^{1}$}

\affiliation{$^{1}$\mbox{Department of Physics and Astronomy G. Galilei,~University of Padova,}\\  INFN and CNISM, via Marzolo 8, I-35151 Padova, Italy\\
$^{2}$\mbox{Laboratory of Ecohydrology, School of Architecture, Civil and Environmental Engineering,} \\ \mbox{\'Ecole Polytechnique F\'ed\'erale de Lausanne, CH-1015 Lausanne, Switzerland}\\
$^{3}$\mbox{Department of Aquatic Ecology,~Eawag, CH-8600 D\"ubendorf, Switzerland}\\
$^{4}$\mbox{Department of Applied Mathematics,~University of Leeds, Leeds, LS2 9JT, UK} \\
$^5$\mbox{Department ICEA, University of Padova,~I-35131 Padova, Italy}\\}


\begin{abstract}
The motility of microorganisms in liquid media is an important issue in active matter and it is not yet fully understood. Previous theoretical approaches dealing with the microscopic description of microbial movement have modeled the propelling force exerted by the organism as a Gaussian white noise term in the equation of motion. We present experimental results for ciliates of the genus \textit{Colpidium}, which do not agree with the Gaussian white noise hypothesis. We propose a new stochastic model that goes beyond such assumption and displays good agreement with the experimental statistics of motion, such as velocity distribution and velocity autocorrelation.

\end{abstract}

\maketitle
If we could look at what happens in a puddle with a microscope we would probably see a myriad of different microorganisms moving around. Bacteria and protists of different species, with body size spanning from 1$\mu$m to 100$\mu$m and which move at speeds of less than 1mm per second. Most of these living creatures do not just diffuse passively but actively swim around looking for nutrients or better living conditions, employing different propulsion mechanisms: some are covered in beating cilia, some have a few rotating flagella, others crawl by changing the shape of their body. Their movement has many interesting features and has been studied from various viewpoints \cite{bray,Berg93}, at different levels of description.\\

The approach we choose in this work is that of neglecting the molecular and biophysical basis of motion peculiar to each organism (bacteria, ciliates, amoeba,..) and focus instead on its statistical properties. Thus, we aim to develop a stochastic approach for microbial mobility that is meant to describe a large class of microorganisms, independently of the particular means of propulsion they might use. This approach, which may appear like an overly simplified description of a complex phenomenon, could be an appropriate starting point to shed light on the key features of microbial behavior.\\

Previous microscopic approaches describing the motion of self-propelled particles in viscous fluids in the absence of external signals employed various modification of the Ornstein-Uhlenbeck (O-U) process
\begin{align}
\frac{dx}{dt}&=v\\
\frac{dv}{dt}&=-\rho v + \xi(t),
\end{align}
where $\xi(t)$ is a Gaussian white noise. As a general framework, the O-U process proved apt to describe some key features of microbial motion, for instance the experimentally observed persistence of velocity \cite{persistence, mendez}.
However, the O-U process fails to recover other statistical properties of the motion, in particular the probability distribution of swimming speed P($v$). In fact the O-U process predicts a Gaussian-distributed velocity, while observations of various species highlighted strong non-Gaussian features \cite{mendez,persistence,exponential,hydra}.
Moreover, it is unclear how the O-U process should be extended in two or three dimensions. For instance, the simplest 2D extension 
\begin{align}
\label{eq:O-U}
\frac{dv_x}{dt}&=-\rho v_x + \xi_x(t)\\
\frac{dv_y}{dt}&=-\rho v_y + \xi_y(t),
\end{align}
where $\xi_x(t)$ and $\xi_y(t)$ are two independent white noise terms, makes the unrealistic assumption that the processes along the two axes are independent. Other models have been suggested to avoid such an assumption. An excellent review can be found in \cite{mendez}.\\ 


We propose a model equation that is inspired by the O-U process, but where the propulsion force exerted by the organism is represented not by a Gaussian white noise but by a suitable stochastic process $\boldsymbol\sigma_t$. The properties of $\boldsymbol\sigma_t$ are prompted by experimental observation of swimming trajectories of the ciliate \textit{Colpidium}. Experimental data were used to fit the proposed model equation. The proposed $\boldsymbol\sigma_t$ is non-Gaussian, and its components along orthogonal axes are not independent. The P($v$) predicted by the model is, consequently, non-Gaussian, and gives a satisfactory approximation to the measured speed distribution.\\

\begin{figure}[tb]
\centering
\begin{tabular}{cc}
\subfloat[a][$P(|\vec{v}|)$]
{\includegraphics[width=.5\columnwidth]{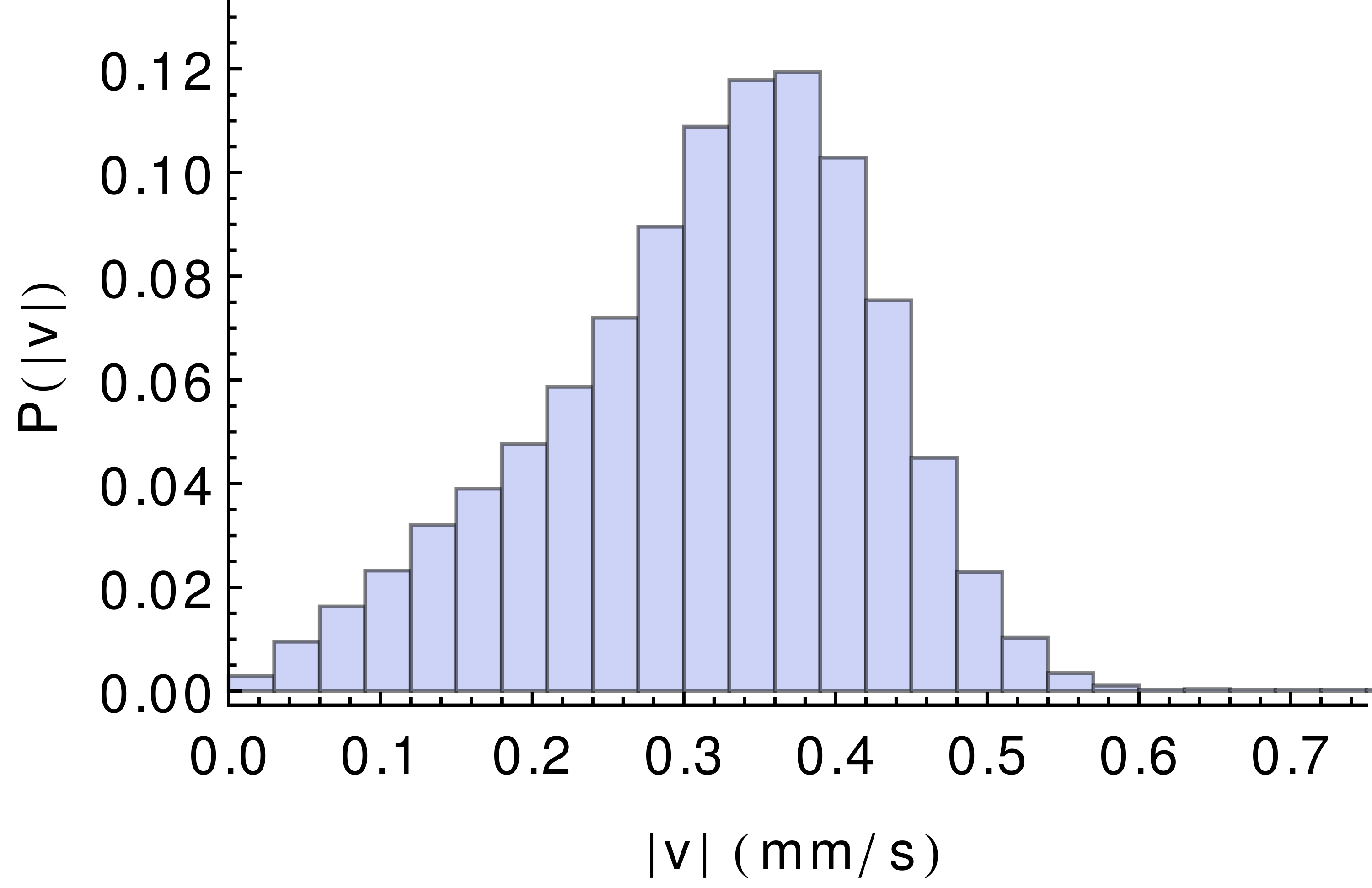}} &
\subfloat[b][$P(v_x,0)$]
{\includegraphics[width=.5\columnwidth]{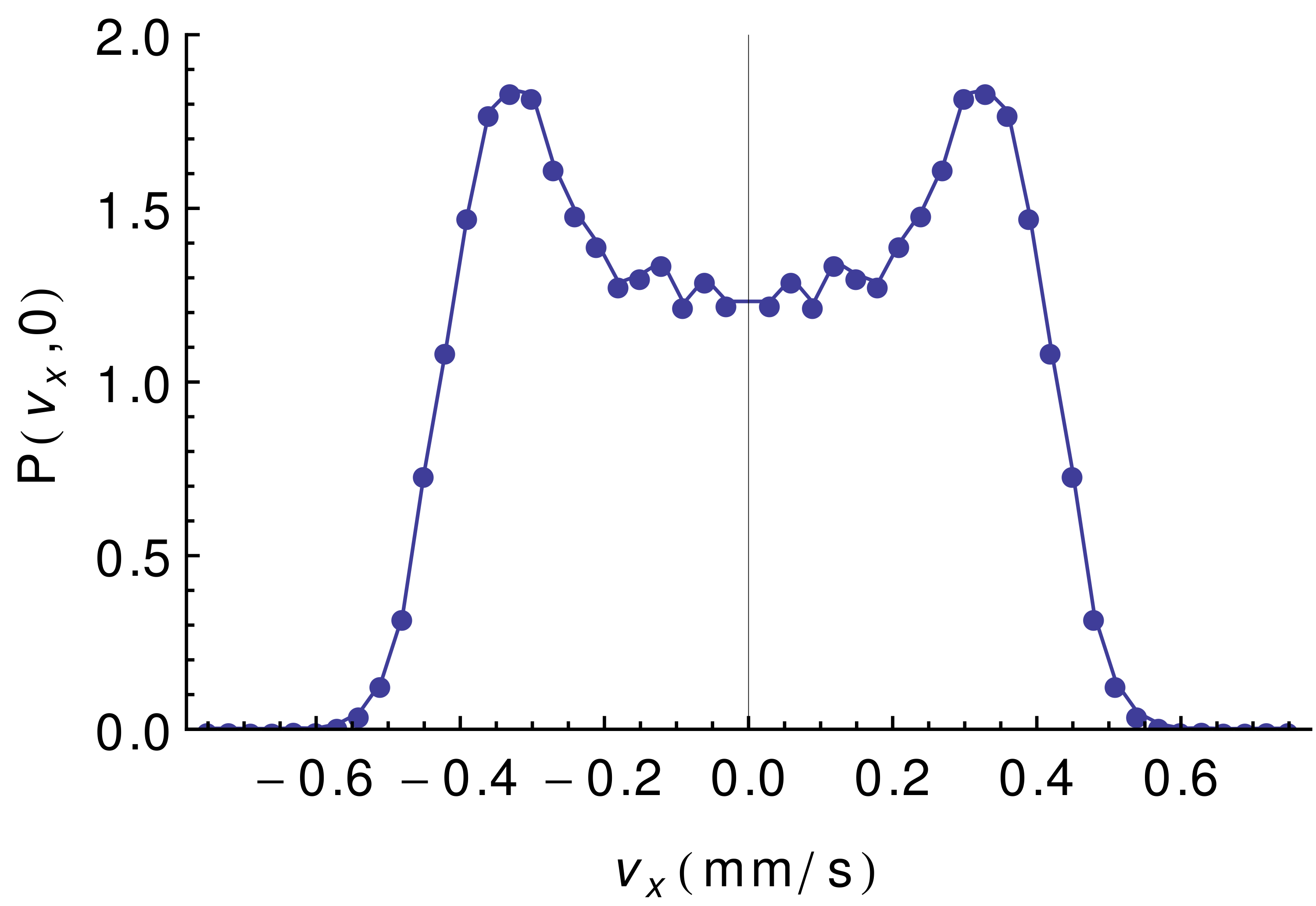}}\\
\end{tabular}
\\
\subfloat[c][C(t)]
{\includegraphics[width=.5\columnwidth]{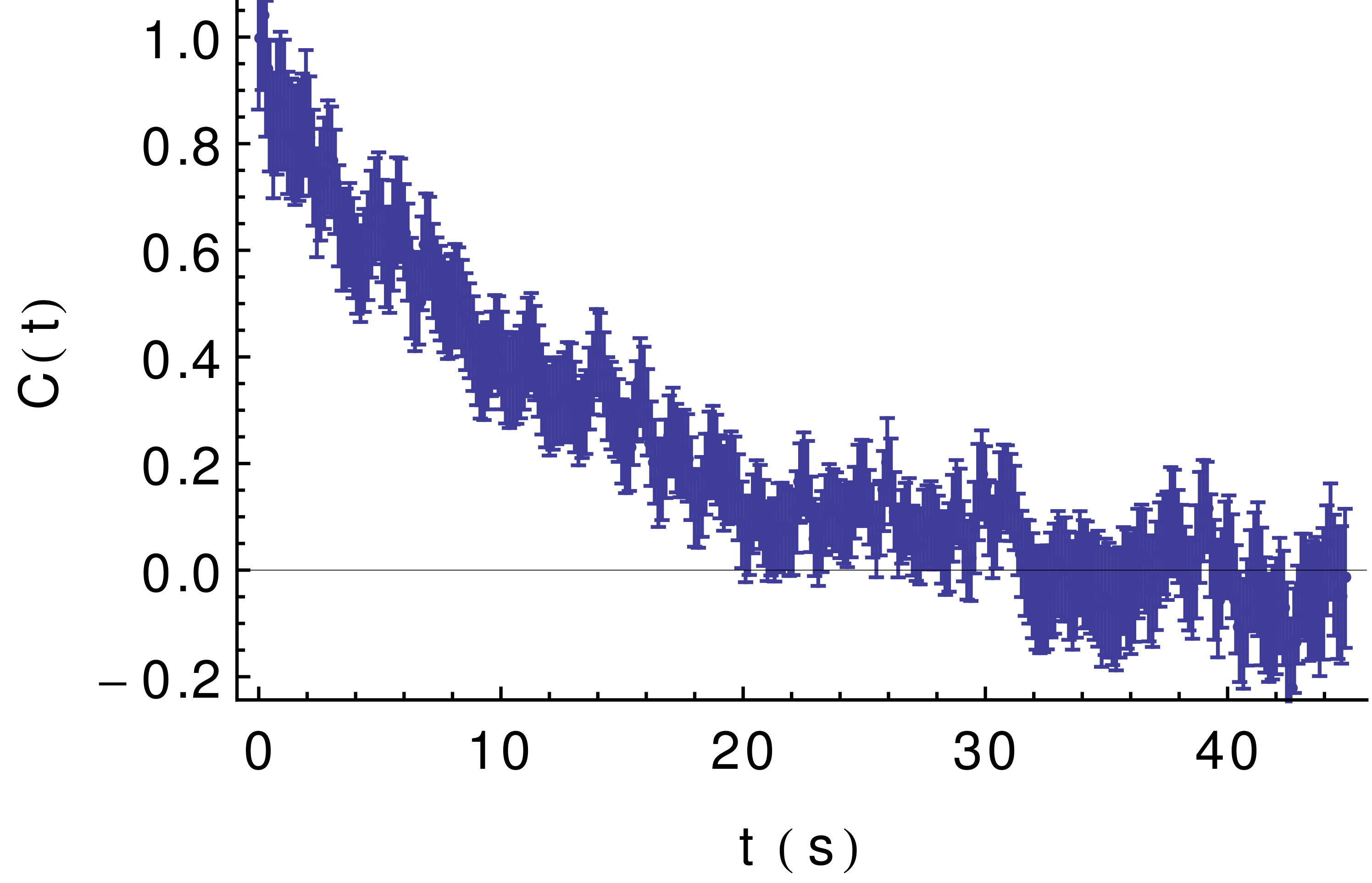}}
\caption{Experimental data for: (a) velocity's moduli distribution; (b) velocity's distribution $P(v_x,0)$; (c) velocity autocorrelation.}
\label{fig:data}
\end{figure}

\begin{figure}
\centering
\subfloat[a][]
{\includegraphics[width=0.6\columnwidth]{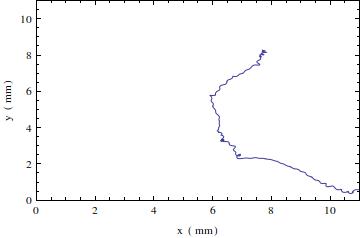}}
\\
\subfloat[b][]
{\includegraphics[width=0.6\columnwidth]{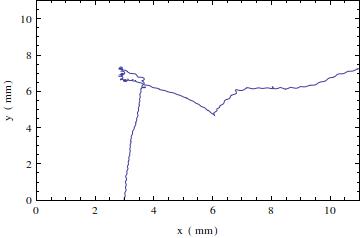}}
\caption {Two examples of reconstructed trajectories of \textit{Colpidium} individuals.}
\label{fig:trajec_data}
\end{figure}

\textit{The experiment} 
The study organism, \textit{Colpidium} sp., was acquired at Carolina Biological Supply (USA). The culture medium was made of local spring water and Protozoan Pellets (Carolina Biological Supply) at a density of $0.45$g/L. The viscosity of the medium can be assumed to be equal to the viscosity of water. Three bacteria species (\textit{Serratia fonticola}, \textit{Breviacillus brevis}, \textit{Bacillus subtilis}) were added to the culture and acted as a food source for \textit{Colpidium} sp. The culture of \textit{Colpidium} sp. was initialized two weeks before the experiment and kept under fluorescent light at a constant temperature of 22$^\circ$C. The experiment was performed by introducing $1$ mL of \textit{Colpidium} sp. in a Sedgewick Rafter Counting Cell S52 (Pyser - SGI, UK) made of glass. The counting cell was placed under the objective of a stereomicroscope and a video was recorded at a framerate of $0.15$s$^{-1}$; the visible area in the video was $11,0$x$8,3$mm. The experiment was repeated three times by extracting $1$mL of culture for each measurement. Videos were analyzed with the software \textit{Mathematica}, version 9.0, which was used to extract the position of each individual at all times. The plugin \textit{MOSAIC} \cite{sbalzarini05} for the software \textit{ImageJ} was used to reconstruct trajectories (see figure \ref{fig:trajec_data}). A total of $1035$ trajectories were recorded, each lasting between $30$s and $60$s (such variability is due to organisms exiting the observation window). Ciliates are observed to perform helical trajectories, which is attributable to the different relative orientation of consecutive rows of cilia along the cell surface \cite{acta_protozoologica}. Microbial motion in the experiment can be considered two dimensional, being the vertical component of the motion negligible, as results from a visual inspection.
We computed the mean velocity between two measures of the position as $\vec{v}(t)=\frac{\vec{x}(t+\Delta t)-\vec{x}(t)}{\Delta t}$, using $\Delta t$=0.45s. This is approximately the duration of one of spiral of the motion, and was chosen in order not to follow too closely the helical motion. In fact, since we do not aim at describing the spiraling pattern of \textit{Colpidium}, but the motion of a large class of organisms, it is deemed reasonable that we overlook such specific details. 
The velocity distribution $P(\vec{v})$ exhibit a volcano-like shape peaked at speeds of approximately $0.4$mm/s. The distribution appears rotationally invariant, as expected from the lack of any directional stimulus in the environment. Thus, we restrict our investigation to the distribution of moduli $P(v)$ (fig. \ref{fig:data} (a)) instead of the full two dimensional distribution.
From the distribution of moduli we obtained a slice of the rotational-invariant distribution $P(v_x,v_y)$, which allows visual inspection of the volcano-like shape and facilitates comparison between the model and the data (figure \ref{fig:data}).

The velocity autocorrelation
\begin{equation}
C(t)=\frac{\langle \vec{v}(t) \cdot \vec{v}(0) \rangle }{ \langle \vec{v}(0)^2\rangle}
\end{equation}
was also computed from the experimental data and is shown in figure \ref{fig:data} (c).\\

\begin{figure}
\includegraphics[width=\columnwidth]{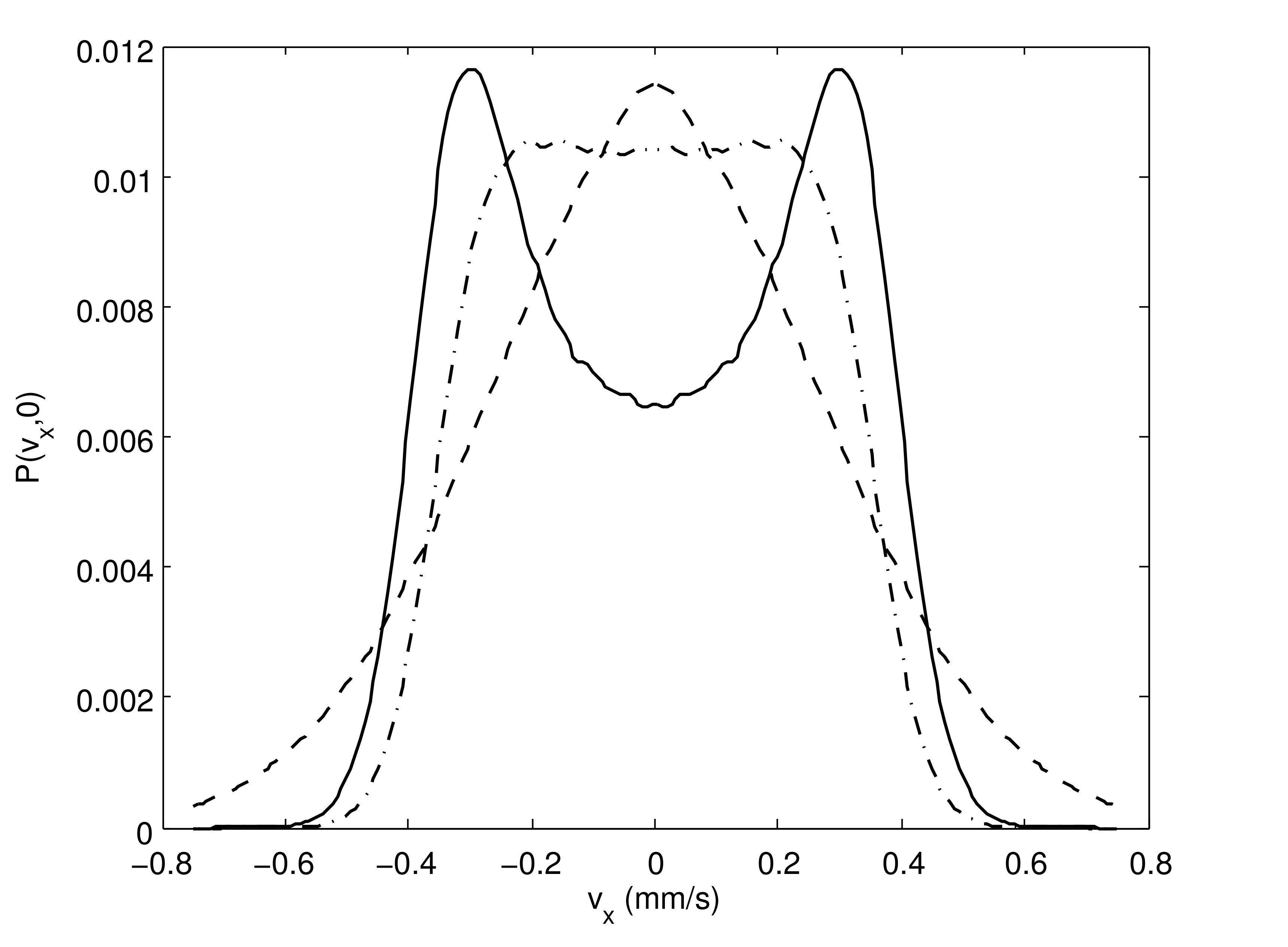}
\caption {\small{$P(\vec{v})$ in the two different regimes of the model and at the crossover between the two regimes:  $\rho \tau >1$ (\textit{solid line}),  $\rho \tau=1$ (\textit{dash-dot line}),  $\rho \tau <1$ (\textit{dashed line}).}}
\label{fig:regimi}
\end{figure}

\textit{The model} We propose the following generalization of equations \eqref{eq:O-U} and (4):
\begin{equation}
\label{eq:model}
\dot{\vec{v}}=-\rho \vec{v}+\vec{\sigma},
\end{equation}
where $\vec{\sigma}$ represents the self-propulsion force, which is modeled as a stochastic process. Microbial motion displays directional persistence, alternating runs in a given direction at an about constant speed to changes of direction  (in figure \ref{fig:trajec_data} two examples of the experimental trajectories). This motion is typically referred to as ``run and tumble''. In the case of \textit{Colpidium}, though, tumbles are instantaneous, differently from what is observed in other species.
Consequently, 
our hypothesis for the force is that of a Markovian process $\vec{\sigma_t}=\sigma_t (\cos \theta_t,\sin \theta_t)$ with infinitesimal generator 
\begin{small}
\begin{multline}
\label{eq:gen_force}
\mathcal{L} f(\theta, \sigma)=\\
=\frac{1}{2 \pi \tau} \int_0^{2\pi} \dd \theta ' \int_{-\infty}^{+\infty} \dd \sigma ' \frac{1}{\sqrt{2\pi} \xi} e^{-\frac{(\sigma '-a)^2}{2 \xi^2}}[f(\theta ',\sigma ')-f(\theta,\sigma)]
\end{multline}
\end{small}
that is, jumps are exponentially distributed with rate $1/\tau$ and each jump is associated to the choice of both a new direction (an angle $\theta$ uniformly distributed in $[0,2\pi]$) and an updated modulus for the propelling force, $\sigma$, where $\sigma\sim$N(a,$\xi$). Note that this distribution allows in principle also negative values of $\sigma$, but for the values of $a$ and $\xi$ found fitting the experimental data the probability of negative values is negligible so we did not encounter any problem in the analysis.\\
The average time of permanence of the process in a certain state is $\tau$. Thus, the stochastic force exhibits no directional preference (because we are modeling movement in the absence of external stimuli) and a modulus that fluctuates around a non-zero mean.
Having established the properties of the stochastic force, eq. \eqref{eq:model} defines the velocity process, $\vec{v}_t$.  We may picture this process in the following way: if at time t the stochastic force has a value $\vec{\sigma}_t$, the velocity will approach the limit velocity $\vec{v}_l=\vec{\sigma}_t/\rho$ with a characteristic time $\tau_c=1/\rho$.  With probability $e^{-t'/ \tau}dt'$ a time in the interval ($t'$, $t'+dt'$] is chosen and a new value of the noise, $\vec{\sigma}_{t+t'}$ is drawn and the velocity will be pulled towards a new limit velocity.\\
The process ($\vec{v}_t$,$\theta_t$,$\sigma_t$) is Markovian, and the equation for the stationary joint probability $\rm P$($\vec{v}$,$\theta$,$\sigma$) can be written as
\begin{footnotesize}
\begin{multline}
\label{eq:stat_prob}
-\nabla_{\vec{v}}[(-\rho \vec{v}+\vec{\sigma}) p(\vec{v},\theta,\sigma)]+\\
+\frac{1}{2 \pi \tau} \int_0^{2\pi} d\theta ' \int_{-\infty}^{+\infty} d\sigma ' \frac{1}{\sqrt{2\pi} \xi} e^{-\frac{(\sigma '-a)^2}{2 \xi^2}}[p(\vec{v},\theta ',\sigma ')-p(\vec{v},\theta,\sigma)]\\
=0
\end{multline}
\end{footnotesize}
Equation \eqref{eq:stat_prob} is, to the best of our knowledge, not solvable analytically.  Instead, we can study the stationary velocity distribution for the process $\vec{v}_t$, $P(\vec{v})$, with the help of numerical simulations.\\
The equation of motion, eq.\eqref{eq:model}, is used used to simulate the process $\vec{v}_t$ by adopting a Gillespie algorithm \cite{gillespie}, where the timing of direction changes are extracted from an exponential distribution of rate $1/\tau$ and $v_x$ and $v_y$ are updated using the analytical solution of eq. \eqref{eq:model} between two successive direction updates. The values of $\vec{v}$ are then sampled every $\Delta t$
and are used to construct the theoretical stationary distribution $P(\vec{v})$. 
Simulations show that $P(\vec{v})$ is strongly affected by the relative values of $\rho$ and $\tau$ and exhibits a crossover between two regimes: if i) $\rho \tau >1$, the limit velocity $\vec{v}_{l}=\frac{\vec{\sigma}}{\rho}$ is reached before the stochastic force changes; if ii) $\rho \tau <1$, the characteristic time of direction update is too small to allow complete relaxation. In the case i) $P(\vec{v})$ displays the volcano-like shape observed experimentally, while in the case ii) $P(\vec{v})$ is peaked in $\vec{v}=0$ and we may suppose that it tends to a Gaussian in the limit $\tau \rightarrow 0$, $a \rightarrow +\infty$, $\xi \rightarrow 0$ (see figure \ref{fig:regimi}). In this limit, in fact, $\boldsymbol\sigma_t$ tends to a Gaussian white noise. This can be rigorously demonstrated for the 1D case, where $\boldsymbol\sigma_t$ reduces to a Markovian dichotomous noise \cite{lefever,dicotomous}. 
Depending on the study species, regime i) or ii) might be of interest.

\begin{figure}[h]
\centering
\includegraphics[width=\columnwidth]{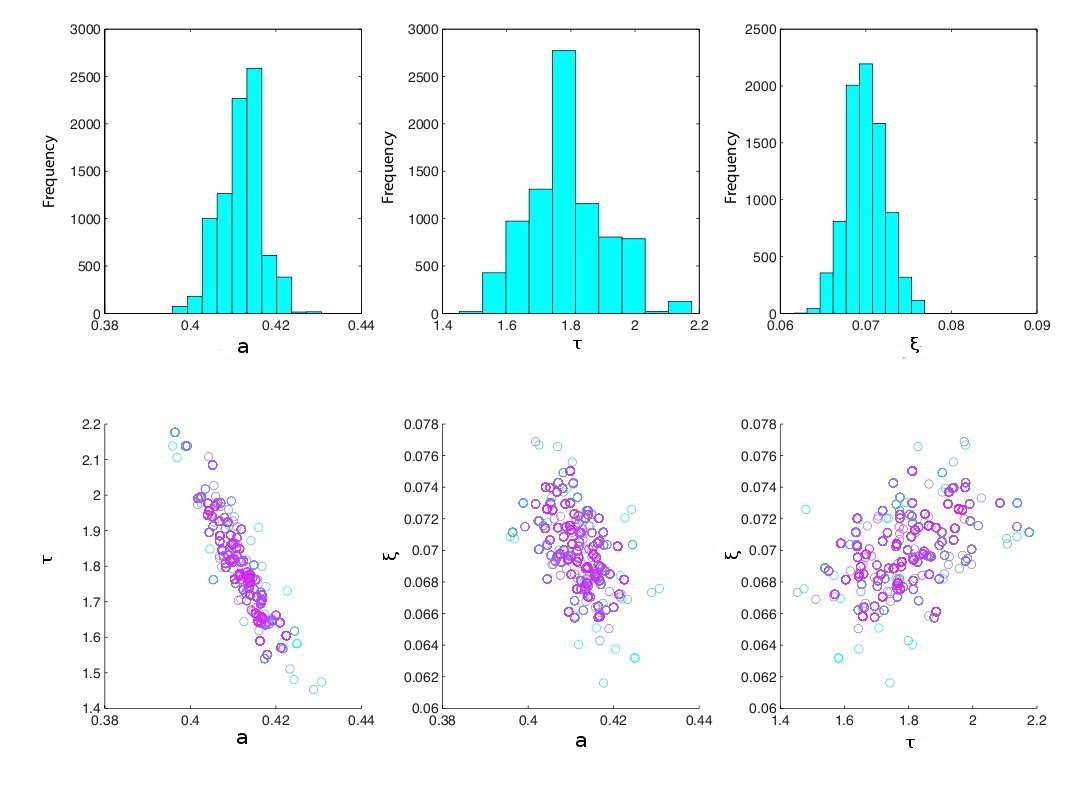}
\caption {Top: Posterior marginal probability distributions for the parameters, ensuing from the MCMC algorithm, with $\rho$ set to $\rho=1$. Bottom: Scatter plots of the parameters explored after reaching stationarity. The color indicates the log-likelihood for each choice of the parameters (light blue = low, purple = high).}
\label{fig:posteriors}
\end{figure}

\textit{Comparison between model and data}  As mentioned above, our model predicts a volcano-shaped velocity distribution P($v$) in certain parameter ranges. This allows to perform a quantitative comparison between model and data and to find the set parameters that best describes the experimentally observed motion. The comparison of the velocity distribution between model and data, however, only allows fitting three of the four parameters ($\rho$, $\tau$, $a$, and $\xi$) of the model. In fact, a rescaling of time in the model equation \eqref{eq:model} shows that the stationary distribution $P(\vec{v})$ only depends on $a/\rho$, $\xi /\rho$, $\tau \rho$. To fit all four parameters, we first fit $a/\rho$, $\xi /\rho$, $\tau \rho$ to the experimental velocity distribution and subsequently fit $\rho$ to the experimental velocity autocorrelation. The fitting of the velocity distribution was performed via the MCMC algorithm DREAM \cite{DREAM}, which allows computing the posterior distribution of the parameters given the data. Fig. \ref{fig:posteriors} shows such distribution (obtained marginalizing the stationary distribution of the MCMC), while table \ref{tab:best_fit} shows the parameter values that give the maximum likelihood fit. Errors were estimated as the width of the posterior distribution at half height.\\

\begin{table}[h]
\centering
\begin{tabular}{c | c }
\hline
$a$/$\rho$ & 0.41 $\pm$ 0.01 mm/s\\
$\xi$/$\rho$ & 0.070 $\pm$ 0.005 mm/s\\
$\tau \rho$ & 1.7 $\pm$ 0.1\\
\hline
\end{tabular}
\caption{Best fit parameters.}
\label{tab:best_fit}
\end{table}

\begin{figure}[h!]
\begin{center}
\includegraphics[width=0.9\columnwidth]{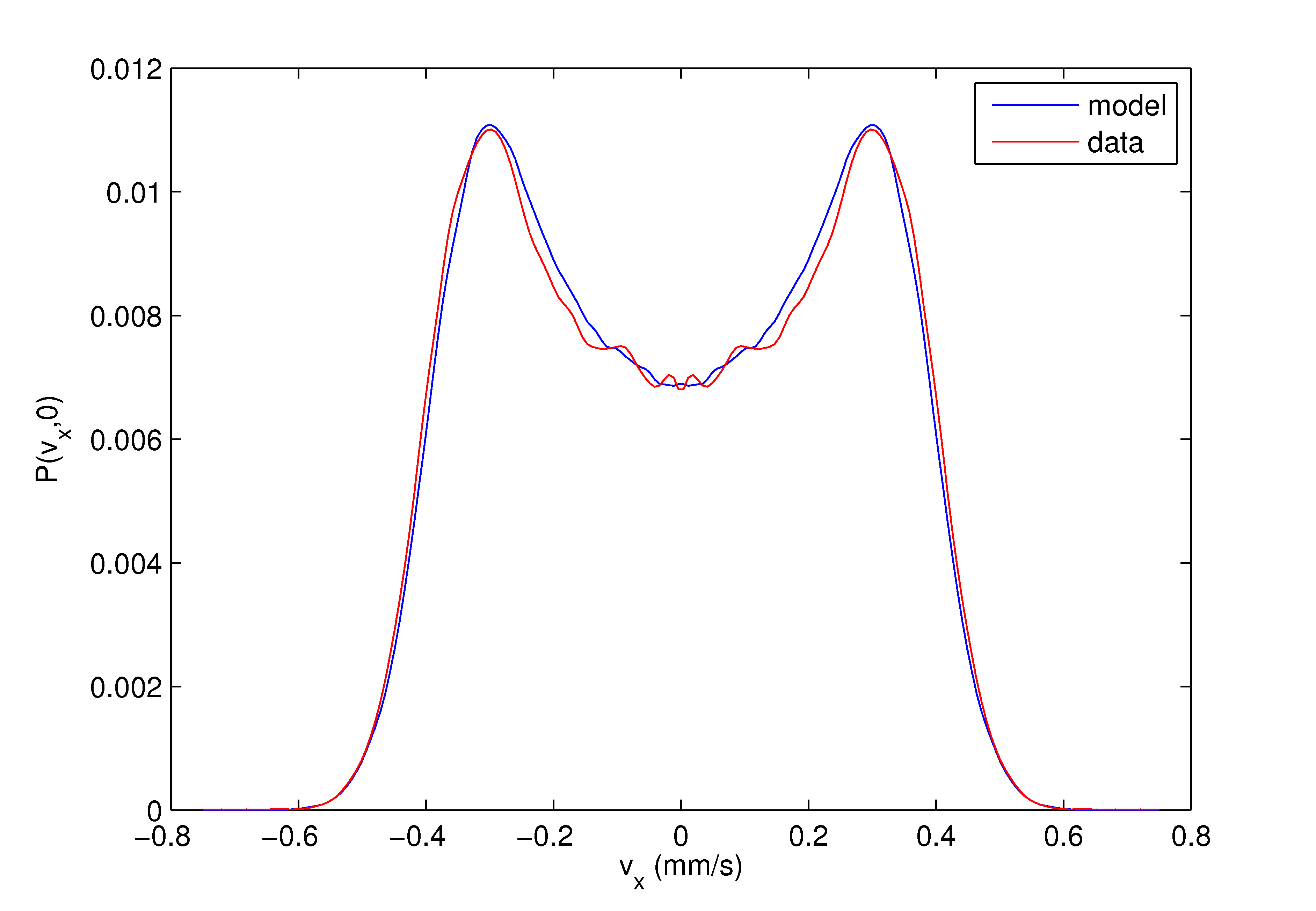}
\end{center}
\caption{Best fit for the stationary velocity distribution $P(v_x,0)$.}
\label{fig:fit_v}
\end{figure}

\begin{figure}[h]
\begin{center}
\includegraphics[width=0.8\columnwidth]{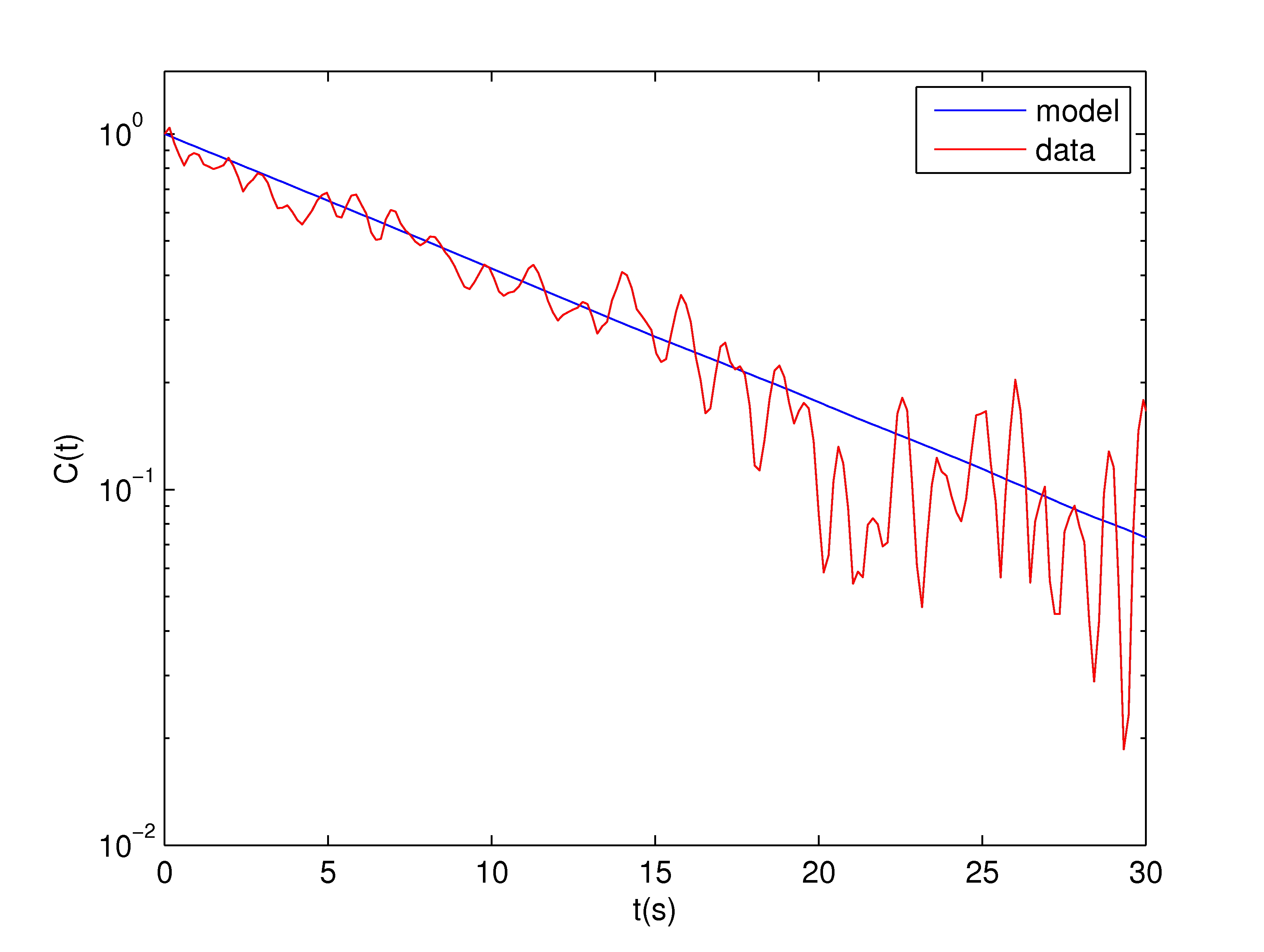}
\caption{\small{Best fit for the velocity autocorrelation. The oscillations in the experimental curve (red) are due to the spiraling pattern of the motion.}}
\label{fig:fit_autocorr}
\end{center}
\end{figure}

\begin{figure}[h]
\begin{center}
\includegraphics[width=0.8\columnwidth]{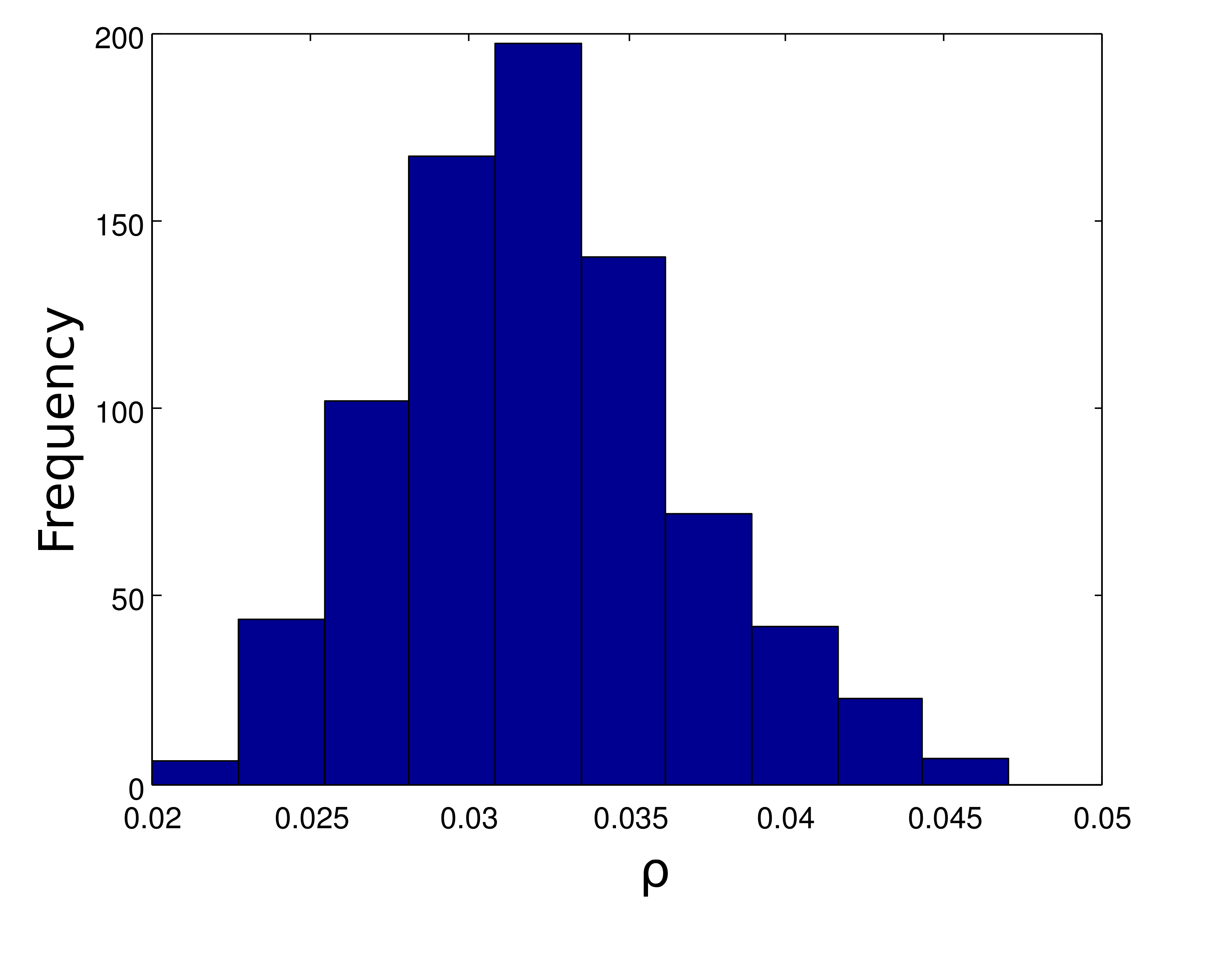}
\caption{\small{Posterior distribution of $\rho$.}}
\label{fig:posterior_ro}
\end{center}
\end{figure}

The resulting best fit for the velocity distribution is shown in figure \ref{fig:fit_v}. The best fit for the autocorrelation was found, instead, by minimization of the $\chi^2$ with a bootstrap method (in figure \ref{fig:posterior_ro} the posterior distribution of the fitted parameter $\rho$). 
Figure \ref{fig:fit_autocorr} shows the best fit of the autocorrelation curve and table \ref{table} reports the estimated values of the parameters.

\begin{table}[h!]
\centering
\begin{tabular}{c | c }
\hline
$\rho$ & 0.033 $\pm$ 0.005 s$^{-1}$\\
$a$ & 0.014 $\pm$ 0.002 mm/s$^2$\\
$\xi$ & 0.0023 $\pm$ 0.0004 mm/s$^2$\\
$\tau$ & 54 $\pm$ 8 s\\
\hline
\end{tabular}
\caption{\small{Best fit parameters.}}
\label{table}
\end{table}
\textit{Conclusions}
With respect to the previous models, the proposed model brings the significant novelty of modeling the force with a process which is non-Gaussian and non-white. These phenomenologically driven hypotheses lead to a good agreement with the data. In particular, the model succeeds in simultaneously fitting the probability distribution of velocities and the autocorrelation.\\
A promising feature of this model is its compatibility with probability distributions of different shapes, depending on the range of parameters. This broadens the set of organisms whose motion the model could describe, including those with a $P(v)$ peaked at the origin, as that observed in \cite{selmeczi, hydra}. To inquire further whether the model is specific for the species \textit{Colpidium} or has a more general validity, in the near future we plan to compare the model predictions with data relative to another species, the rotifer \textit{Cephalodella} sp. A possible future development is that of extending the model to "informed" movement, i.e. movement in presence of external stimuli.

\end{document}